\documentclass[graybox]{svmult}

\usepackage{type1cm}        
                            
\usepackage{makeidx}         
\usepackage{graphicx}        
                             
\usepackage{multicol}        
\usepackage[bottom]{footmisc}

\usepackage{hyperref}        

\usepackage{newtxtext}       
\usepackage[varvw]{newtxmath}

\makeindex

\begin{document}

\title*{Dimensionality reduction using pseudo-Boolean polynomials for cluster analysis}

\author{Tendai Mapungwana Chikake and Boris Goldengorin}

\institute{
Tendai Mapungwana Chikake 
\at 
	Department of Discrete Mathematics, 
	Phystech School of Applied Mathematics and Informatics, 
	Moscow Institute of Physics and Technology, 
	Institutsky lane 9, 
	Dolgoprudny, 
	Moscow region, 141700, Russian Federation
	\email{tendaichikake@phystech.edu}
\and 
Boris Goldengorin \at
	Department of Mathematics, 
	New Uzbekistan University, 
	Tashkent, 
	100007, 
	Uzbekistan.
	The Scientific and Educational Mathematical Center “Sofia Kovalevskaya Northwestern Center for Mathematical Research” in Pskov State University, 
	Sovetskaya Ulitsa, 21, Pskov, Pskovskaya oblast’, 180000, Russian Federation.
	Department of Discrete Mathematics, 
	Moscow Institute of Physics and Technology, 
	Russian Federation, 
	\email{goldengorin.bi@mipt.ru}
}

\maketitle

\abstract*{
	We introduce usage of a reduction property of penalty-based formulation of pseudo-Boolean polynomials as a mechanism for invariant dimensionality reduction in cluster analysis processes. 
	In our experiments, we show that multidimensional data, like 4-dimensional Iris Flower dataset can be reduced to 2-dimensional space while the 30-dimensional Wisconsin Diagnostic Breast Cancer (WDBC) dataset can be reduced to 3-dimensional space, and by searching lines or planes that lie between reduced samples we can extract clusters in a linear and unbiased manner with competitive accuracies, reproducibility and clear interpretation. 
}
\abstract{
	We introduce usage of a reduction property of penalty-based formulation of pseudo-Boolean polynomials as a mechanism for invariant dimensionality reduction in cluster analysis processes. 
	In our experiments, we show that multidimensional data, like 4-dimensional Iris Flower dataset can be reduced to 2-dimensional space while the 30-dimensional Wisconsin Diagnostic Breast Cancer (WDBC) dataset can be reduced to 3-dimensional space, and by searching lines or planes that lie between reduced samples we can extract clusters in a linear and unbiased manner with competitive accuracies, reproducibility and clear interpretation. 
}

\section{Introduction}

In the fields of data visualization and cluster analysis, dimensionality reduction mechanisms play pivotal roles in providing better understanding of relations and clusters in input data. 
Dimensionality reduction techniques work by transforming data from high-dimensional spaces into low-dimensional spaces such that the low-dimensional representations retain meaningful properties of the original data, ideally close to their intrinsic dimensions \cite{scholkopf_nonlinear_1998}.

Real-world data is often available in high-dimensional spaces which are usually cognitively and computationally hard to process \cite{Johnstone}.
Human observers as well as presentation mediums readily available, like 2-dimensional papers or screens, are presented with representational challenges whenever data is available in higher than 3-dimensional spaces and consequently the identity of classes, useful or noisy features becomes harder to discover \cite{Johnstone}. 

Data scientists often spend enormous amounts of time and effort digging for relevant features that determine classes or those features that bring useless noise in datasets. 
Automated systems like artificial neural networks can be used in the feature selection processes \cite{notleyExaminingUseNeural2018a}, but often require large amounts of data and challenges like feature superposition \cite{elhage2022superposition}, underfitting, overfitting, and interpretation concerns arise \cite{burnham_model_2002}.

In fields where large numbers of observations and/or large numbers of variables exist such as signal processing, computer vision, speech recognition, neuroinformatics, and bioinformatics, usage of dimension reduction techniques is crucial \cite{kopp_simultaneous_2022}.
Dimensionality reduction simplifies cluster analysis tasks for both human and machine processors \cite{kopp_simultaneous_2022}.

In this work, we observe that our powerful dimensionality reduction method can assist in reducing the abuse of statistical methods and/or artificial neural networks in tasks that can be solved in combinatorial steps.
We show this by qualifying our dimensionality reduction method, followed by linear clustering of reduced samples.
This advantage is available because our reduction method has invariability properties, and it is intuitively easy to interpret.
The problems of invariability and interpretability are among the top problems of currently available dimensionality reduction methods \cite{sarveniazi_actual_2014}.
Dimensionality reduction tools that lack interpretability and/or invariability can be disfavoured in critical tasks such as clustering models that input medical tests/measurements to predict a diagnosis. 
Our work is directed to solving such challenges.

Cluster analysis or clustering is the task of grouping a set of objects in such a way that objects in the same group (called a cluster) are more similar (in some sense) to each other than to those in other clusters \cite{peterson_overview_2010}. 
Clustering is a major task of exploratory data analysis, and a common technique for statistical data analysis, used in many fields, including pattern recognition, image analysis, information retrieval, bioinformatics, data compression, computer graphics and machine learning \cite{peterson_overview_2010}.

Abuse of statistical methods and/or artificial neural networks often arise in cases where data is minimal.
This abuse may result in overfitted, irreproducible or hard-to-interpret solutions which may be undesirable to use in some critical tasks.

The invariant manipulation based on formulation of pseudo-Boolean polynomials presented in this work, can enable cluster analysts to extract simple rules of associations in data without the need for machine learning.
The formulation of pseudo-Boolean polynomials is very simple, computationally efficient, invariant to ordering, and easy to explain and reproduce.

Our overall contributions in this work are:
\begin{enumerate}
    \item Qualifying the usage of the reduction property of penalty-based pseudo-Boolean polynomials formulation for dimensionality reduction of multidimensional data where it is feasible.
    \item Reducing overdependence on data-driven approaches in solving problems that can be solved with combinatorial steps.
\end{enumerate}

Dimensionality reduction using pseudo-Boolean polynomials formulation, revolves around the manipulation of the \textit{reduction} and \textit{equivalence} properties of penalty-based pseudo-Boolean polynomials \cite{goldengorin_cell_2013}.

We present our results on classical Wisconsin Diagnostic Breast Cancer (WDBC) \cite{street_nuclear_1993} and Iris Flower datasets \cite{fisher1936use}, which have too few samples, such that the usage of a data-driven methods like artificial neural networks for clustering would result in abuse.

The Iris Flower dataset \cite{fisher1936use} has samples of size ${1 \times 4}$ and present challenges of identifying clusters by plotting on a Cartesian plane for the analyst while the Wisconsin Diagnostic Breast Cancer (WDBC) \cite{street_nuclear_1993} dataset, has samples of size ${1 \times 30}$ that result in 30-Dimensionality representation that would be incomprehensible for Cartesian plot based analysis.

Our proposed method is limited to data whose samples can be represented as cost matrices where each cell represents a cost relationship of its respective column and row.
In the experiment sections, we show how these complex datasets can be reduced to ${2 \times 1}$ and ${3 \times 1}$ dimensionality which are easily analysed on a Cartesian plane and Cartesian space respectively.
Simple linear demarcations are then used to qualify the label of a given sample without any machine learning process or non-linear alteration of data.

\section{Related work}

Principal Component Analysis (PCA) and the T-distribute Stochastic Neighbour Embedding (t-SNE) are arguably the most popular dimensionality reduction methods in cluster analysis tasks. The choice of usage of either, is usually case based. 
The T-distribute Stochastic Neighbour Embedding is often used for visualizing high dimensional data. It works by converting similarities between data points to joint probabilities and tries to minimize the Kullback-Leibler divergence between the joint probabilities of the low-dimensional embedding and the high-dimensional data \cite{vanderMaaten}. 
T-SNE has a cost function that is not convex, i.e., with different initializations different reductions may result \cite{vanderMaaten}. 
The non-convex nature of the cost function in the t-SNE tool is a major drawback in comparison with the Principal Component Analysis method as well as the pseudo-Boolean polynomials-based reduction that we qualify in this paper. 

Dimensionality reduction by pseudo-Boolean polynomials formulation ensures unique reduced Hammer-Beresnev polynomials, regardless of possible difference in ordering of input matrices \cite{goldengorin_cell_2013}.
Consequently, our method is guaranteed to output the same reductions, regardless of difference in initializations or input orderings.

Principal Component Analysis (PCA), is an orthogonal linear transformation that transforms data to a new coordinate system such that the greatest variance by some scalar projection of the data comes to lie on the first coordinate (called the first principal component), the second-greatest variance on the second coordinate, and so on \cite{jolliffe_ian}.
The major drawback of PCA is that it changes the distances involved in our data because it reduces dimensions in a way that preserves large pairwise distance better than small pairwise distance \cite{jiang_bias_2000}.
These changes can be very sensitive to our algorithms, especially when working with Euclidean distance-based algorithms.

Modifications to the Principal Component Analysis (PCA) method exists like the kernel principal component analysis (kernel PCA) \cite{scholkopf_nonlinear_1998} and Graph-based kernel PCA \cite{bengio_nonlocal_2006}, that can be employed in a nonlinear manner but the distance and data dependence concerns prevail.

Other techniques include Linear discriminant analysis (LDA) \cite{mclachlan_discriminant_1992}, and Uniform manifold approximation and projection (UMAP) \cite{mcinnes_umap_2020}.
All of these techniques, raise all or some of the concerns outlined above as they are largely dependent on data distribution to operate.

The pseudo-Boolean polynomials approach presented in this work, operates in isolation for each individual sample, such that no distribution biases are introduced to any given sample. 
The combinatorial operations are constant and invariant to ordering across all samples.

\section{Methods} 

In mathematics and optimization, a pseudo-Boolean function is a function of the form ${f: \mathbf{B^n} \to \mathbb{R}}$, where ${\mathbf{B} = \{0, 1\}}$ is a Boolean domain and ${n}$ is a non-negative integer called the degree of the function.  \cite{boros_pseudo-boolean_2002}

We utilise a penalty based formulation of pseudo-Boolean polynomials described in \cite{albdaiwi_data_2011} for data aggregation which provides us an invariant dimensionality reduction property that we present in this work. 

\cite{goldengorin_cell_2013} highlights fundamental reduction properties of pseudo-Boolean polynomials which guarantee the maintenance of the underlying initial information while reducing the size of the problem. 

We extend the objective goal in \cite{albdaiwi_data_2011} by describing our problem as a problem of minimising the cost of describing a given sample and thereby reducing the dimensionality of the particular sample. 

Given a sample where observables (e.g. physical quantities, types of measurements, e.t.c) ${I = \{1, 2, ..., m\} }$ (ordered by their correlation strengths to labels), their physical measurements ${\textit{J} = \{1, 2, ..., n\} }$, and ${p}$ the maximum dimensionality size desired, to minimise the cost of describing the given sample, our task is to find a set ${S \subseteq I}$  with  ${|S| = p}$ minimising the prespecified objective function.
 
We define the instance of the problem by an ${m \times n}$ matrix ${\textit{C} = [c_{i  j}] }$ of costs (distances, bandwidth, time, (dis)similarities, (in)significance, etc.), ${\textit{j} \in \textit{J} }$ and  ${\textit{i} \in \textit{I} }$, and the goal is to find a set ${\textit{S} \subseteq \textit{I} }$ with ${|S| = p}$, such that we minimise total cost 
\begin{equation}
	\label{eqn:p_1}
	f_c(S) = \sum_{j \in J}  min\{\mathrm{c_{ij}} | i \in S\},
\end{equation} with the assumption that entries of ${C}$ are non-negative and finite \cite{albdaiwi_data_2011}.

According to \cite{albdaiwi_data_2011}, the objective function ${f_c(S)}$ in (\ref{eqn:p_1}) is a kind of problem that can be formulated in terms of pseudo-Boolean polynomials and from \cite{boros_pseudo-boolean_2002} all pseudo-Boolean polynomials can be uniquely represented as multilinear polynomials of the form 

\begin{equation}
	\label{eqn:mlp}
	f(\mathbf{y}) = \sum_{S \subseteq I}  \mathrm{c_S} \prod_{i \in S} \mathrm{y_i}
\end{equation}

Pseudo-Boolean polynomials formulation is achievable in polynomial time and allows us to achieve compact representations of relatively large problems \cite{albdaiwi_data_2011}.

From (\ref{eqn:mlp}), ${\prod_{i \in S} y_i }$ is the term of the monomial  ${ c_S \prod_{i \in S} y_i}$. Monomials with the same term are called similar monomials \cite{goldengorin_cell_2013}, and they can be added together in a process called \textit{reduction} \cite{goldengorin_cell_2013} which is central to our dimensionality reduction solution as it allows us to reduce the number of columns in the initial cost matrix.

In addition to compacting large problems, there exists different instances that have similar (reduced) Hammer–Beresnev polynomials, mainly because similar monomials can be aggregated and disaggregated \cite{goldengorin_cell_2013}. 

These properties are essential in cluster analysis because samples which might look dissimilar in higher dimensional space, can actually converge to similarity in their reduced pseudo-Boolean polynomials form.

This work seeks to exploit these fundamental properties: representational reduction and \textit{equivalence} as dimensionality reduction and clustering mechanisms respectively.

By treating measurable attributes of multidimensional data, like physical measurements, pixel positioning and intensity distribution in image data, and other describable/measurable attributes as \textit{information costs} of describing  samples, we can formulate for each sample, a cost matrix ${C}$ which we can manipulate and reduce, in an ordering invariant manner, by pseudo-Boolean polynomials formulation and achieve lower dimension representation of each sample independent of any other samples in the dataset.

The \textit{equivalence} \cite{goldengorin_cell_2013} property and other distance comparisons can then be applied on the reduced data representation for cluster analysis. 

Visualizing a matrix sample of size ${1 \times n}$ where ${n \in \{1, 2, 3\}}$ is easily comprehensive by scatter plotting along 1-D, 2-D, and 3-D planes respectively. 
If classes are present in the data, we can identify linear or non-linear lines or plane separators that demarcate boundaries of clusters in the data.

The pseudo-Boolean approach to dimensionality reduction in measured features for sample clustering is a penalty-based approach that relies on the fact that we require attributes that positively distinguish underlying classes for each instance to be sufficiently represented based on their importance to the classifier. 

The task seeks to minimize measurements that \textit{insignificantly} contribute to the identity of a sample in a specific class. 

A sample is described by an ${m \times n}$ matrix ${C = [c_{i,j}]}$ where ${I}$ represents the measured feature while ${J}$ represent the measurement such that columns of the matrix contain homogenous quantities. 
In reducing the dimensionality of samples in the Iris Flower dataset \cite{fisher1936use} for example, ${I}$ represents measured features, \textit{sepal, and petal} while ${J}$ represent the measurements \textit{width and length} thereof.

We define the decisive insignificance ${S}$ of an attribute to classifying a sample into a specific cluster as ${f_c(S)}$ in (\ref{eqn:p_1})
and the dimensionality reduction task is the problem of finding 
\begin{equation}
	\label{eqn:m_2}
	S^* \in  arg\,  min\{f_c(S) : \emptyset \subset S \subseteq I, |S| = p \},
\end{equation}
where ${p}$ is the output dimension size, which we however choose to be ${|J|}$ such that no measurement is lost.

By processing our samples into pseudo-Boolean polynomials we achieve the sample representations with the least possible information costs needed to represent them.

\section{Experimental Setup}

\subsection{The Iris Flower dataset}

Table \ref{Tab:Tcr} shows the Iris Flower dataset \cite{fisher1936use}, a classical and popular dataset in the machine learning community. 
The task on this dataset is to classify Iris plants into three species \textit{(Iris setosa, Iris versicolor, and Iris virginica)} using the lengths and widths of their petals and sepals. The dataset contains 150 samples.

Our method requires that data is structured as matrices, where columns are measurements and rows are the features measured. 
For this dimensionality reduction task, we first reshape the structure of all instances from ${1 \times 4}$ sized instances to ${2 \times 2}$ sized instances, where rows represent the measurement type \textit{(Sepal, Petal)} and the columns represent the measurements \textit{(Length and Width)} of the features collected from 3 different species:

\begin{table}
    \caption{Iris Dataset\label{Tab:Tcr}}{}
    \begin{tabular}{lrrrrl}
		ID &  sepal length (cm) &  sepal width (cm) & petal length (cm) & petal width (cm) & target \\
		
		1  &                5.4 &               3.0 &                4.5 &               1.5 &  versicolor \\
		2  &                7.7 &               2.8 &                6.7 &               2.0 &   virginica \\
		3  &                5.2 &               3.4 &                1.4 &               0.2 &      setosa \\
		4  &                4.8 &               3.4 &                1.9 &               0.2 &      setosa \\
		.  &                  . &                 . &                  . &                 . &           . \\
		.  &                  . &                 . &                  . &                 . &           . \\
		.  &                  . &                 . &                  . &                 . &           . \\
		150  &              6.6 &               3.0 &                4.4 &               1.4 &  versicolor \\

    \end{tabular}
\end{table}

We transform the samples into costs matrices by making rows of measured features (Sepal, Petal) and columns of measurements (Length, Width) resulting in 2x2 costs matrices. We show a transformed sample in Table \ref{Tab:Tcr_restructured}

\begin{table}
\caption{Transformed instance\label{Tab:Tcr_restructured}}{}
\begin{tabular}{lrrrrl}

{} &  length (cm) &  width (cm) \\

sepal &                5.4 &               3.0 &                 \\
petal &                4.5 &               1.5 &                 \\

\end{tabular}
\end{table}

\begin{figure}
    \centering
    \includegraphics[width=0.9\textwidth]{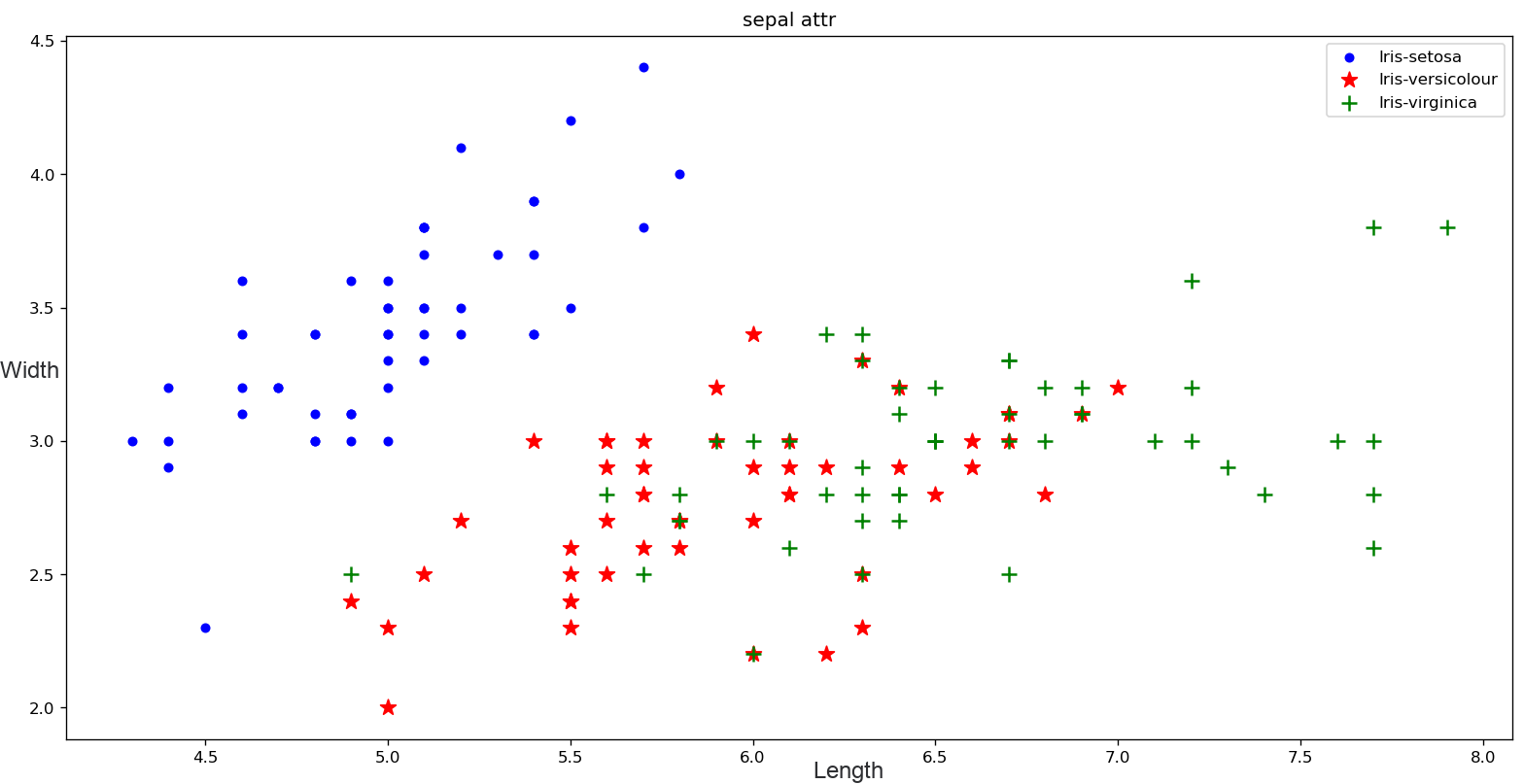}
    \caption{Iris dataset scatter plot on sepal length-width\label{fig:scatter_sepal}}{}
\end{figure}

\begin{figure}
    \centering
    \includegraphics[width=0.9\textwidth]{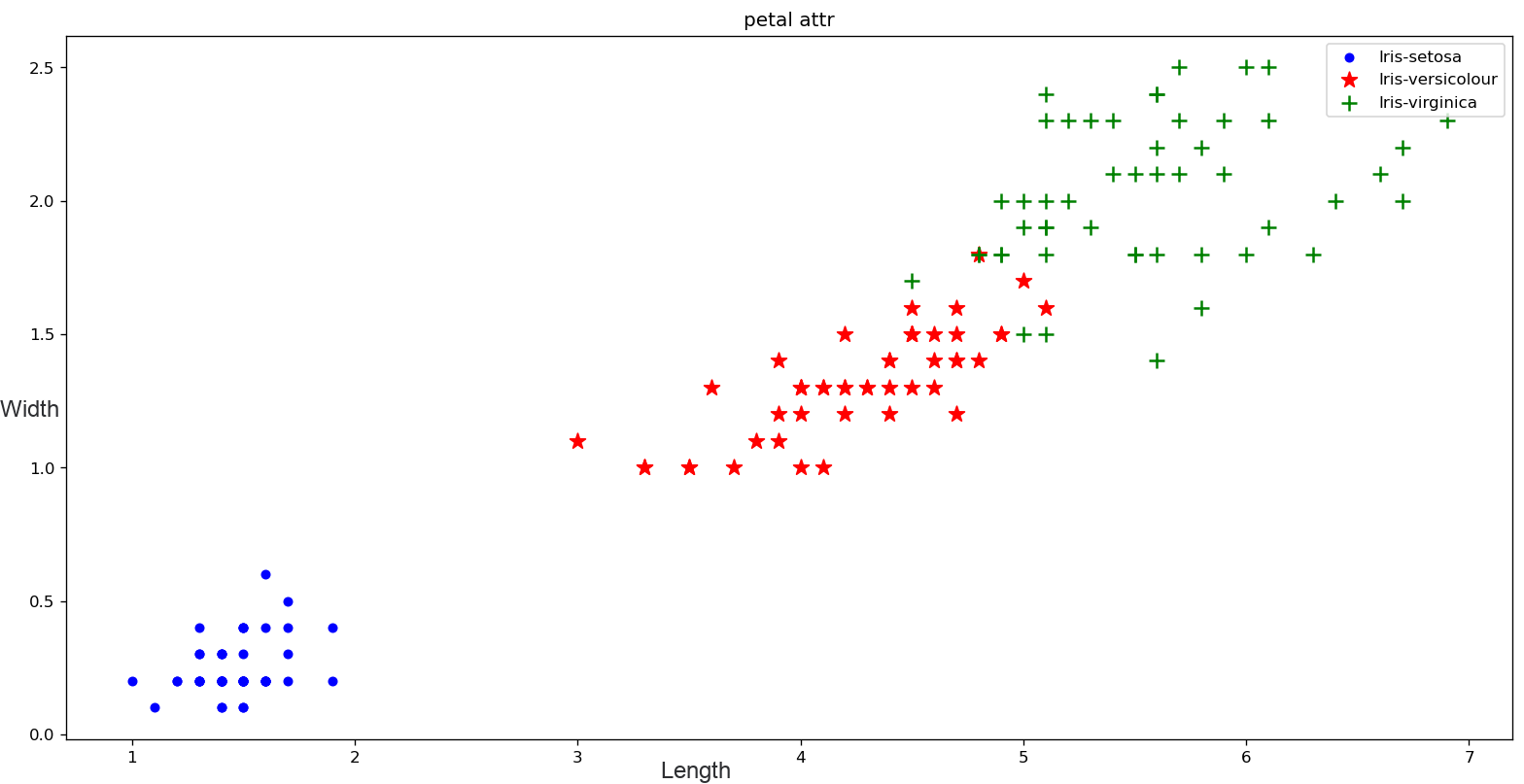}
    \caption{Iris dataset scatter plot on petal length-width\label{fig:scatter_petal}}{}
\end{figure}

One cluster in the Iris Flower dataset, \text{Iris-Setosa} is linearly separable from others while \textit{Iris-virginica} and \textit{Iris-versicolour} are not linearly separable between each other as shown in Fig. \ref{fig:scatter_sepal} and Fig. \ref{fig:scatter_petal}

Fig. \ref{fig:scatter_sepal} shows the scatter plot on lengths and widths of sepal measurements while Fig. \ref{fig:scatter_petal} shows the scatter plot on lengths and widths of petal measurements of the samples.

By analysing the scatter plots, it is visible that one cluster in the Iris Flower dataset \textit{Iris-Setosa} is linearly separable from others while the other two, \textit{Iris-virginica and Iris-versicolour} are not linearly separable between each other as shown in Fig. \ref{fig:scatter_sepal} and Fig. \ref{fig:scatter_petal}, 

Applying the pseudo-Boolean polynomials reduction program, we reduce all samples to ${2 \times 1}$ matrices. 
After applying combinations of like terms and dropping of zero columns, the pseudo-Boolean polynomials of the instance in Table \ref{Tab:Tcr_restructured} reduces to
\[
\begin{bmatrix}
 6.0\\
 2.4y_2\\
\end{bmatrix}
\]

Applying the reduction method on every other sample, results in a list of samples reduced to ${2 \times 1}$ matrices of the form ${a + by_2}$ which can be plotted and clustered on a Cartesian plane by simple boundary lines. 

\subsection{The Wisconsin Diagnostic Breast Cancer (WDBC) dataset}

The Wisconsin Diagnostic Breast Cancer (WDBC) \cite{street_nuclear_1993} dataset, is another classical and popular dataset in the machine learning community.
The dataset has 569 instances of 30 real-valued features that describe characteristics of the cell nuclei present in digitized images extracted by a fine needle aspirate (FNA) on a breast mass \cite{street_nuclear_1993}.

Ten real-valued features are computed for each cell nucleus:
\begin{enumerate}
	\item radius (mean of distances from centre to points on the perimeter)
	\item texture (standard deviation of greyscale values)
	\item perimeter
	\item area
	\item smoothness (local variation in radius lengths)
	\item compactness (${perimeter^2 / area - 1.0}$)
	\item concavity (severity of concave portions of the contour)
	\item concave points (number of concave portions of the contour)
	\item symmetry
	\item fractal dimension (``coastline approximation'' - 1)
\end{enumerate}

The \textit{mean, standard error (se), and "worst"} or largest (mean of the three largest values) of these features are computed for each image, resulting in 30 features \cite{street_nuclear_1993}.

The task on this dataset is to predict whether a breast cancer diagnosis is \textit{benign} or \textit{malignant} based on these features. 

The best known predictive accuracy (97.5\%) was obtained by using a separating plane in the 3-D space of Worst Area, Worst Smoothness and Mean Texture features using repeated 10-fold cross-validations, and this classifier has correctly diagnosed 176 consecutive new patients as of November 1995 \cite{street_nuclear_1993}. 

The separating plane was obtained using Multisurface Method-Tree (MSM-T) \cite{bennett_decision_1992}, a classification method which uses linear programming to construct a decision tree where relevant features are selected using an exhaustive search in the space of 1-4 features and 1-3 separating planes \cite{bennett_decision_1992}.

Samples of size ${3 \times 10}$, result in 30-Dimensionality representations, that are hard to visualise on a Cartesian plot and consequently hard for the analyst to identify the features that are useful in making an accurate diagnosis.

To qualify our dimensionality reduction tool, we input each sample as a ${3 \times n; n \in {1, 2, 3, \dots, 10}}$ matrix where rows ${J}$ represent the measurement type (\textit{Mean}, \textit{Standard error (se)}, \textit{Worst}) and the columns ${I}$ the measured features.

Applying the pseudo-Boolean polynomials formulation, we reduce all samples to ${3 \times 1}$ matrices which can be plotted and clustered on a Cartesian space by a simple boundary plane. 
Since our formulation runs in polynomial time, we can iterate all possible arrangements of measured features and discover the set of measured features which has the best fitting plane that separates samples into \textit{benign} or \textit{malignant}.

\section{Results and Discussion}

\subsection{The Iris Flower dataset}

\begin{figure}
    \includegraphics[width=0.9\textwidth]{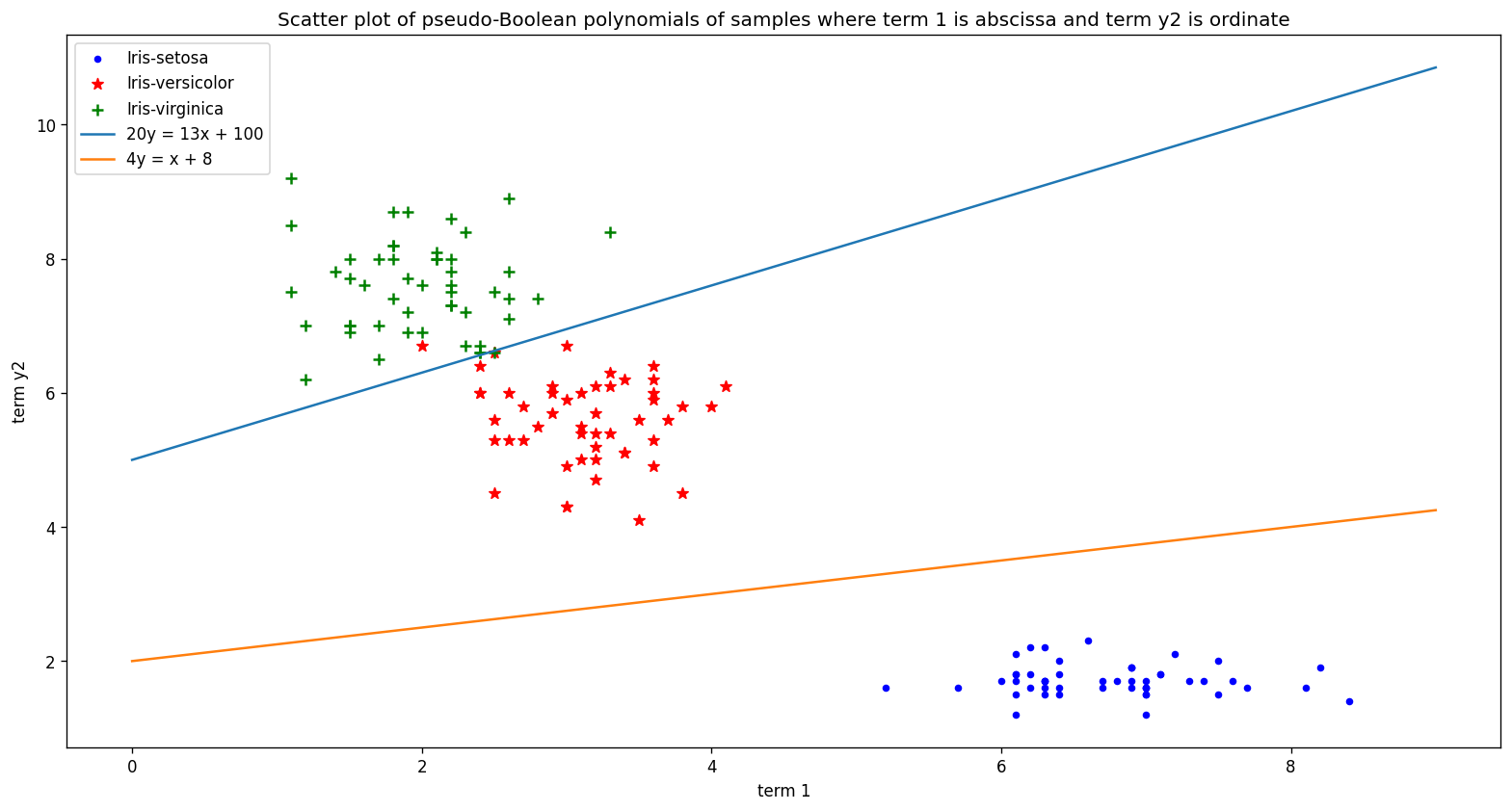}
    \centering
    \caption{Cartesian plot on resultant ${2 \times 1}$ dimensions after reduction by pseudo-Boolean polynomials formulation and the best boundary lines\label{fig:scatter_aggregated}}{}
\end{figure}

Fig. \ref{fig:scatter_aggregated} illustrates the identification of cluster boundaries by simple identification of lines that best separate the aggregated samples on a Cartesian plane. 
Plotting the coefficients of the reduced instances, where the terms 1 and y2 are abscissa and ordinate respectively, allows us to visualize properly the possible clusters from the data.
Aggregated widths are represented by constant terms, while the aggregated lengths are represented by linear terms in the resulting pseudo-Boolean polynomials.

The most important thing to note is that this reduction correctly represents the original instances with less information cost, and allows us to discover a pair of straight lines that separate the respective clusters. 

All instances that lie in the region ${y \le \frac{x}{4} + 2}$, are classified safely as \textit{Iris-setosa}, while the line ${y = \frac{13x}{20} + 5}$, separates \textit{Iris-versicolor} from \textit{Iris-virginica}.

It is also important to note that, some instances which seemed distinct of each other, actually have similar reduced pseudo-Boolean polynomials form. e.g. 

\[
\begin{bmatrix}
 5.4 & 3.4\\
 1.7 & 0.2\\
\end{bmatrix}
\]
and 

\[
\begin{bmatrix}
 5.1 & 3.7\\
 1.5 & 0.4\\
\end{bmatrix}
\]
all reduce to 

\[
\begin{bmatrix}
 1.9 \\
 6.9 y_2
\end{bmatrix}
\]
and in perfect confirmation of the \textit{equivalence} condition; the instances also lie in the same cluster. 

Looking at the single outlier,
\[
    \begin{bmatrix}
     5.1 & 3.7\\
     1.5 & 0.4\\
    \end{bmatrix}
\]
which was reduced to 
\[
\begin{bmatrix}
 2.0 \\
 6.7 y_2
\end{bmatrix}
\]
and classified as \textit{Iris-virginica} instead of \textit{Iris-versicolor} we observe the overfitting flaw of learning-based cluster methods such as support-vector machines and \cite{sarkar_xbnet_2022}'s X-Boosted artificial neural network, which output 100\% cluster accuracy on some test runs. 
The original measurements of this outlier, perfectly fit the cluster \textit{Iris-virginica} as there are samples of \textit{Iris-virginica} that have very little difference in measured values to this outlier, while its values are also significantly distinct from the other \textit{Iris-versicolor} samples.
The incorrectly clustered instance might be attributed to misclassification by the persons who labelled the dataset, or perhaps just a naturally occurring outlier. 

This finding exposes, the overfitting concerns that arise from learning-based methods, like the Space Vector Machine (SVM) and \cite{sarkar_xbnet_2022}'s X-Boosted artificial neural network, that would report 100\% accuracy in some tests.
Additionally, when using these other methods, changing the train/test data results in varied accuracies, and consequently exposing the invariability and reproducibility qualities that our method guarantees. 

\subsection{The Wisconsin Diagnostic Breast Cancer (WDBC) dataset}

\begin{figure}
    \includegraphics[width=1.0\textwidth]{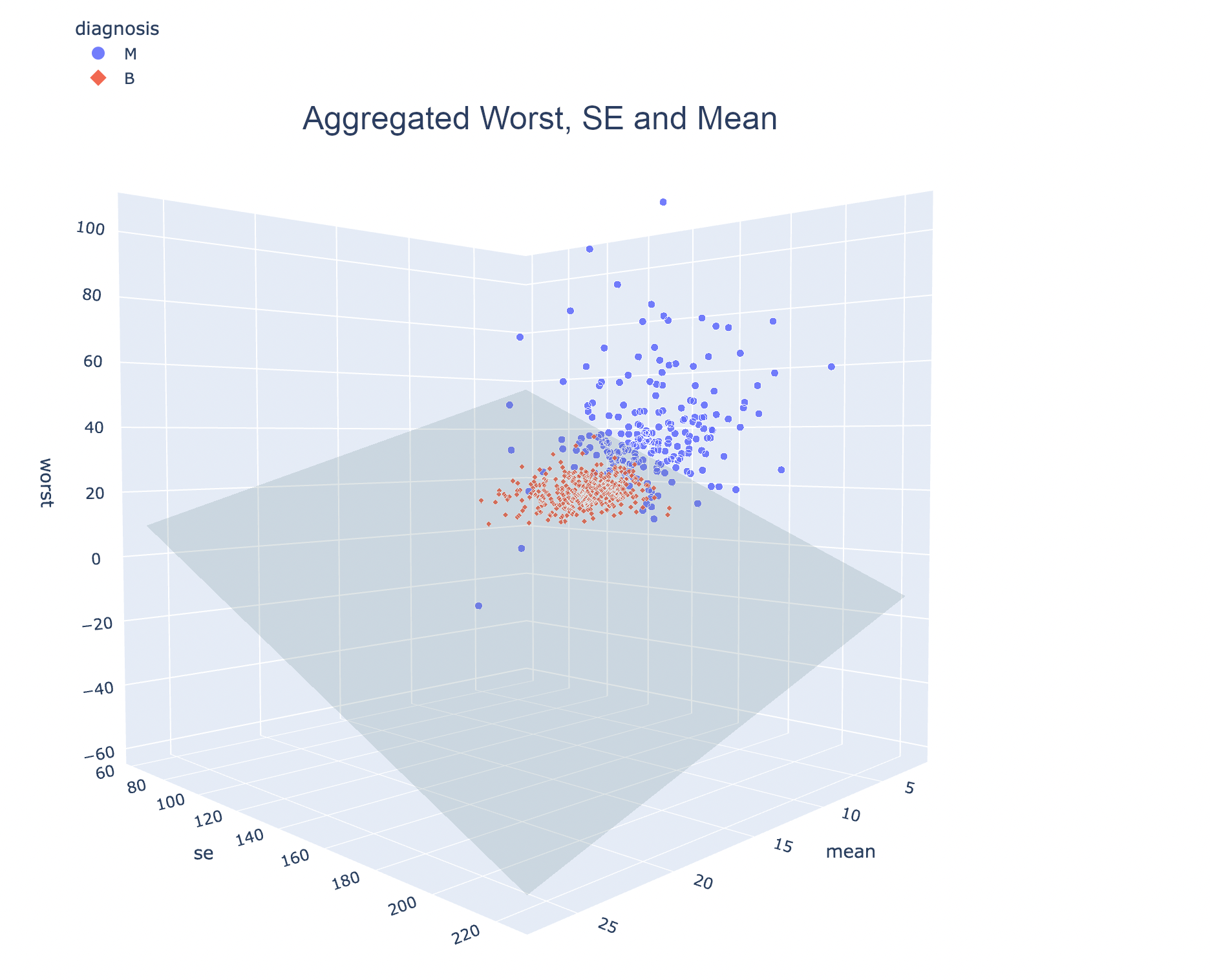}
    \centering
    \caption{Cartesian plot on resultant ${3 \times 1}$ dimensions after reduction by pseudo-Boolean polynomials formulation and the best separating plane\label{fig:scatter_bc}}{}
\end{figure}

Plotting the coefficients of the reduced instances just as in the previous dataset, allows us to visualize properly the possible clusters from the data. Fig. \ref{fig:scatter_bc} shows the cluster plot of the 3-dimensional aggregated measurements in the dataset and the linear plane ${z = 85x - 2y  - 0.4}$ that best separates samples into two classes.

The reduction correctly represents the original instances with less information cost, and allows us to discover a combination of features where a plane that best separates the respective clusters can be discovered. 

Of all the combinations of features explored, the combination with the best separating plane (95.4\% accuracy) consisted of 
\begin{enumerate}
	\item radius
	\item texture
	\item perimeter
	\item smoothness
	\item compactness
	\item concavity
	\item symmetry
	\item fractal dimension
\end{enumerate}
excluding \textit{area} and \textit{concave points} features.

Instances with the shortlisted features are separable by a plane ${z = 85x - 2y  - 0.4}$ at an accuracy of 95.4\%.

Although, falling short of the accuracy(97.5\%) reported in \cite{street_nuclear_1993}, our method manages to present the dimensionality reduction capacity of a simple and invariant method that is solely based on manipulation of orderings. 

Additionally, the orthogonal distance of a sample from the separating plane, serves as the confidence of the clustering plane. The higher the distance, the more confident we are about the clustering accuracy and vice-versa.

As shown in the experimental results reported in this work, our method can be used for unsupervised clustering of multidimensional data as well as in feature selection processes in cluster analysis.
Our method allows us to have a better understanding of how multidimensional features contribute to classification of samples in an invariant and explainable manner and in some cases achieve unbiased and unsupervised clustering in cluster analysis processes.

\section{Conclusion}

In this paper, we managed to showcase a combinatorial method for dimensionality reduction for cluster analysis, based on the formulation and reduction of pseudo-Boolean polynomials. 

We tested our method on simple datasets, and managed to show that we can classify data samples with competitive accuracies by simple and linear data slices (lines and planes).
Our proposed solution is invariant and interpretable while avoiding biases that may be involved when we use statistical methods because each sample is reduced in an independent manner, solely based on its own description.

It can be noted that dimension reduction using pseudo-Boolean polynomials on high-level features is a powerful tool for lossless dimension reduction and has potential of accelerating low memory representation of complex data that empowers other complex tasks like interpretable unsupervised clustering in computer vision, bioinformatics, natural language processing and machine learning.

We expect to reproduce even better state-of-the-art accuracies on other data science tasks, when we apply more powerful tools like decision trees and artificial neural networks on instances in their reduced pseudo-Boolean polynomials forms.

The biggest takeaway from our findings is the invariability and interpretability nature of the dimension reduction process using pseudo-Boolean polynomials formulation.

\section{Acknowledgements}

Tendai Mapungwana Chikake and Boris Goldengorin’s research was supported by Russian Science Foundation project No. 21-71-30005.
 
Boris Goldengorin acknowledges Scientific and Educational Mathematical Center “Sofia Kovalevskaya Northwestern Center for Mathematical Research” for financial support of the present study (agreement No 075-02-2023-937, 16.02.2023)  

\bibliographystyle{spmpsci_unsort}
\bibliography{./bib}

\end{document}